\input{aipcheck.tex}

  \def\selectedoptions{final}

\documentclass[
   \selectedoptions
  ]
  {aipproc}

\layoutstyle{6x9}

\SetInternalRegister\hbadness{8000} 

\newcommand\doingARLO[2][]{%
  \ifx\mmref\undefined #1\else #2\fi
}
  
\begin{document}

%
%
\def\bsig{\mbox{\boldmath $\sigma$}}                          
\def\bsig{\mbox{\boldmath $\Sigma$}}
\def\bgam{\mbox{\boldmath $\gamma$}}
\def\bgam{\mbox{\boldmath $\Gamma$}}
\def\bphi{\mbox{\boldmath $\phi$}}
\def\bphi{\mbox{\boldmath $\Phi$}}
\def\btau{\mbox{\boldmath $\tau$}}
\def\btau{\mbox{\boldmath $\Tau$}}
\def\btau{\mbox{\boldmath $\partial$}}
\def\Delc{{\Delta}_{\circ}}
\def\bp{\mid {\bf p} \mid}
\def\al{\alpha}
\def\bet{\beta}
\def\gam{\gamma}
\def\del{\delta}
\def\Del{\Delta}
\def\te{\theta}
\def\nua{{\nu}_{\alpha}}
\def\nui{{\nu}_i}
\def\nuj{{\nu}_j}
\def\nue{{\nu}_e}
\def\num{{\nu}_{\mu}}
\def\nut{{\nu}_{\tau}}
\def\2te{2{\theta}}
\def\chic#1{{\scriptscriptstyle #1}}
\def\chicl{{\chic L}}
\def\lam{\lambda}
\def\SU{SU(2)_{\chic L} \otimes U(1)_{\chic Y}}
\def\Lam{\Lambda}
\def\sig{\sigma}
\def\'#1{\ifx#1i\accent19\i\else\accent19#1\fi}
\def\O{\Omega}
\def\o{\omega}
\def\s{\sigma}
\def\D{\Delta}
\def\d{\delta}
\def\df{\rm d}
\def\8{\infty}
\def\ld{\lambda}
\def\eps{\epsilon}
\def\ref#1{$^{#1}$}
\def\chicl{{\chic L}}
\def\lam{\lambda}
\def\SU{SU(2)_{\chic L} \otimes U(1)_{\chic Y}}
\def\Lam{\Lambda}
\def\sig{\sigma}
\def\'#1{\ifx#1i\accent19\i\else\accent19#1\fi}
\def\O{\Omega}
\def\o{\omega}
\def\s{\sigma}
\def\D{\Delta}
\def\d{\delta}
\def\df{\rm d}
\def\8{\infty}
\def\ld{\lambda}
\def\eps{\epsilon}
\def\ref#1{$^{#1}$}

\title 
      []
      {Dirac Neutrino Anapole Moment}

\classification{13.15.+g, 13.10.+q, 13.40.Gp, 25.30.Pt}
\keywords{anapole moment of neutrino; charge radius of neutrino; 
electromagnetic properties of neutrino}

\author{L. G. Cabral-Rosetti$^a$, M. Moreno$^b$ and A. Rosado$^{b,c}$}
{address={$^a${\it Instituto de Ciencias Nucleares, Universidad Nacional 
Autónoma de México (ICN-UNAM),\\
Circuito Exterior, C.U., Apartado Postal 70-543, 94510 México, D.F.,  
México.}\\
{$^b${\it Instituto de Física, Universidad Nacional Autónoma de México, 
(IF-UNAM).\\
Circuito de la Investigación Científica, C.U., Apartado Postal 
20-364, 01000 México, D.F.,  México.}\\
$^c${\it Instituto de Física, Benemérita Universidad Autónoma 
de Puebla, (IF-BUAP).\\
Apartado Postal J-48, Colonia San Manuel, Puebla, Puebla. 72570, México.}
}\\
},
  email={luis@nuclecu.unam.mx, matias@fenix.ifisicacu.unam.mx, 
rosado@fenix.ifisicacu.unam.mx},
  thanks={}
}

\copyrightyear  {2001}

\begin{abstract}
We calculate the Dirac neutrino anapole moment ($a_{\nu_l}$) in the context 
of the Standard Model (SM) making use of the Dirac form factor 
$F_{\chic D}(q^2)$ introduced recently by J. Bernabéu, L. G. Cabral-Rosetti,
J. Papavassiliou y J. Vidal by using the Pinch Technique (PT) 
formalism, working in two different gauge-fixing schemes ($R_\xi$ guage and 
the electroweak BFM), at the one loop level. We show that the neutrino 
anapole form factor $F_{\chic A}(q^2)$ and Dirac form factor 
$F_{\chic D}(q^2)$ are related as follows: $F_{\chic A}(q^2) = 
\frac{1}{q^2} F_{\chic D}(q^2)$. Hence, the Dirac neutrino charge radius 
$\langle r^2_{\nu_l} \rangle$ and the anapole moment satisfy the simple 
relation $a_{\nu_l} = \frac{1}{6} \langle r^2_{\nu_l} \rangle$. 
Therefore, we show that the anapole moment (as the charge radius) of 
the neutrino is a physical quantity, which only gets contribution from 
the proper neutrino electromagnetic vertex (in electroweak BFM), and that 
$a_{\nu_l}$ is of the order $10^{-34} \, cm^2$.\ \footnote{Contribution to
the Proceedings of the {\it VIII Mexican Workshop on Particles and Fields} 
of the {\it Divisi\'on de Part{\'\i}culas y Campos de la Sociedad Mexicana de 
F{\'\i}sica} ({\bf DPyC-SMF}), Cd. de Zacatecas, Zacatecas, M\'exico 
November 14-20, 2001.}
\end{abstract}

\date{\today}

\maketitle

\section{Introduction}
In 1987, M. Abak and C. 
Aydin \cite{abak} calculated $a_{\nu_l}$ in the context of the Standard 
Model of the electroweak interaction (SM) \cite{weinberg}, using
the 't Hooft-Feynman gauge \cite{hooft} and conclude that $a_{\nu_l}$ is
too small to be measured. In 1992, A. Góngora-T and R. G. Stuart, 
defined the charge radius and anapole moment of a free fermion as being 
its vector and axial-vector contact interactions with an external 
electromagnetic current and they got, at one loop in the SM, a finite 
and gauge invariant expression for these quantities \cite{gongora}. In 
1987, H. Czyz {\it et al.} \cite{czyz} discussed the anapole moment of 
charged leptons in the context of the SM and showed that this quantity 
is gauge dependent and therefore is not a physical quantity in this 
model. In 1991, M. J. Musolf and B. R. Holstein analized the neutrino
matrix element of the electromagnetic current for low $q^2$ and showed
that the anapole moment is just $\frac{1}{6}$ times the charge radius
\cite{musolf}. In 2000, A. Rosado presented a calculation of the neutrino 
anapole moment in the linear $R_\xi$ gauge at the one loop level in the 
context of the SM and showed explicitly that it is an infinite and gauge 
dependent quantity in this model. He also introduced, the electroweak 
anapole moment $a_{\nu_l}^{\chic {EW}}$ through the elastic scattering
$\nu_l l'$ which is finite, gauge independent, and independent of the 
lepton $l'$ used to define it \cite{rosado1, rosado2}.

In this work, we calculate the neutrino anapole moment $a_{\nu_l}$ in 
the context of the SM of the electroweak interactions $\SU$ making use 
of the Dirac form factor introduced by J. Bernabéu, L. G. Cabral-Rosetti, 
J. Papavassiliou y J. Vidal \cite{bernabeu}. This electromagnetic form 
factor was obtained through the process 
$e^+ e^- \rightarrow \nu_l \bar{\nu_l}$ at the one loop level, 
by using the Pinch Technique (PT) formalism \cite{JMC, CJM, JPP, DAS, wat}, 
working in two different gauge-fixing schemes ($R_\xi$ guage and the 
electroweak BFM), at the one loop level, in the context of SM and becomes 
for arbitrary momentum 
transfer $q^2$: finite, independent on the gauge-fixing parameter, on 
the gauge-fixing scheme employed, on the Higgs and quark sector of the 
theory, and on the properties of the charged lepton used to define it. 
This paper is organized as follows. In the next section,  we show that the
neutrino anapole form factor $F_A(q^2)$ and $F_D(q^2)$ are related as 
follows: $F_{\chic A}(q^2) = \frac{1}{q^2} F_{\chic D}(q^2)$. Hence, the 
neutrino charge radius $\langle r^2_{\nu_l} \rangle$ and the anapole 
moment satisfy the simple relation 
$a_{\nu_l} = \frac{1}{6}\langle r^2_{\nu_l} \rangle$.
Therefore, we show that the anapole moment (as the charge radius) of 
the neutrino is a physical quantity. Also in this section, we give the 
expression of the anapole form factor and the numerical values of the 
anapole moment for the three different neutrino species. Finally, we 
present our conclusions.

\begin{figure}
\includegraphics[height=.19\textheight]{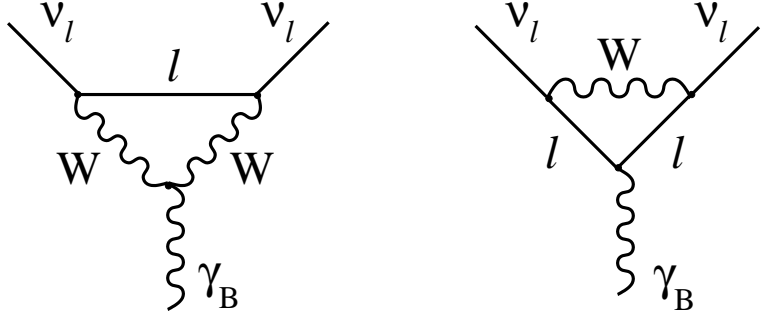}
\caption{
Only two proper vertex diagrams contribute to the neutrino charge radius 
$\langle r^2_{\nu_l} \rangle$ and therefore neutrino anapole form factor 
$F_{\chic A}(q^2)$ at the one-loop level after that the massive gauge 
cancellation, which takes place when the Pinch Technique is used, in the 
electroweak Background Field Method.}
\label{vertex}
\end{figure}

\section{The neutrino anapole moment}

The matrix element of the electromagnetic current in the frame of the SM
can be expressed at the lowest order in $\alpha$, {\it for all} $q^2$, where
$q=p-p'$, in terms of only one form factor $F_{\chic D}(q^2)$ as
\cite{chung, mohapatra, cabral1, C3}

\begin{equation}
{\cal M}_\mu = F_{\chic D}(q^2) \bar u_{\nu_l} (p') \gamma_\mu
(1 - \gamma_5) u_{\nu_l} (p).
\end{equation}

\noindent 
In Fig. 1, we show only the two one-loop diagrams, which contribute to
the Dirac form factor, after that the massive gauge cancellation, which
takes place when the PT is used \cite{bernabeu}, in the electroweak 
Background Field Method \cite{BFM}. For a massless Dirac neutrino we can 
rewrite Eq. (1), for all $q^2$, as follows

\begin{equation}
{\cal M}_\mu = \bar u_{\nu_l} (p') \{ \gamma_\mu f_1(q^2) 
- \gamma_\lambda \gamma_5
[{g^{\lambda}}_{\mu} q^2 - q^{\lambda} q_{\mu}] f_3(q^2) \} u_{\nu_l} (p)
\end{equation}

\noindent 
where\cite{marshak} $f_1(q^2) = F_{\chic D}(q^2)$ and 
$f_3(q^2) = F_{\chic A}(q^2) = \frac{1}{q^2}F_{\chic D}(q^2)$ are the Dirac and anapole form 
factor of the neutrino, respectively, with

\begin{equation}
f_1(0) = F_{\chic D}(0) = 0,
\end{equation}

\begin{equation}
\langle r^2_{\nu_l} \rangle = -\ 6\  \frac{\partial f_1(q^2)}{\partial q^2}
\Bigg |_{q^2=0} \hspace{2mm} = -\ 6\  
\frac{\partial F_{\chic D}(q^2)}{\partial q^2}
\Bigg |_{q^2=0},
\end{equation}

\noindent 
and

\begin{equation}
a_{\nu_l} = f_3(0) = \frac{F_{\chic D}(q^2)}{q^2} \Bigg |_{q^2=0}
\hspace{2mm} = \frac{\partial F_{\chic D}(q^2)}{\partial q^2} 
\Bigg |_{q^2=0}.
\end{equation}

\noindent 
That is,

\begin{equation}
a_{\nu_l} = \frac{1}{6}\ \langle r^2_{\nu_l} \rangle.
\end{equation}

\noindent 
Finally, according to the results given for $F_{\chic D}(q^2)$ in 
Eq. (7.10) of Ref. \cite{bernabeu}, 

\begin{equation}
\begin{array}{c}
\displaystyle
F_{\chic D} (q^2) = -\, \frac{\alpha e}{8 \pi s_{\chic W}^2} 
\Bigg\{1 + \Bigg( \frac{1}{2} + \frac{M^2_{\chic W}}{q^2} \Bigg)
\Big[ B_0 (q^2; m_l^2, m_l^2) 
- B_0 (q^2; M^2_{\chic W}, M_{\chic W}^2) \Big]
\\[0.5cm]
\displaystyle
+\, M_{\chic W}^2 \Bigg( 2 + \frac{M^2_{\chic W}}{q^2} \Bigg)
C_0 (0, q^2, 0; m_l^2, M_{\chic W}^2, M^2_{\chic W})
+\, \frac{(q^2 + M_{\chic W}^2)^2}{q^2}
C_0 (0, q^2, 0; M^2_{\chic W}, m_l^2, m_l^2)
\Bigg\}\, .
\end{array}
\end{equation}

\noindent 
Hence, we conclude that the anapole moment, as the charge radius,
of the neutrino is a physical quantity, which only gets contribution 
from the proper neutrino electromagnetic vertex Fig.~(1). Taking into 
account the relations among scalar, two-points $B_0$ and three-points
$C_0$ Passarino-Veltman functions reported in the Refs. 
\cite{cabral1, C3, cabral2} we get

\begin{equation}
a_{\nu_l} = \frac{G_F}{24 \sqrt{2} \pi^2} \left \{ 3 - 2 \, \log
\Bigg ( \frac{m_l^2}{M_W^2} \Bigg ) \right \},
\end{equation}

\noindent 
The numerical evaluation of the above expression for the three
different neutrino species yields: $a_{\nu_e} = 6.8 \times 10^{-34} \, cm^2$, 
$a_{\nu_\mu} = 4.0 \times 10^{-34} \, cm^2$ and 
$a_{\nu_\tau} = 2.5 \times 10^{-34} \, cm^2$.

\section{Conclusions}

The main goal in this paper has been to calculate the anapole moment of 
the neutrino, making use of the Dirac form factor introduced by J. 
Bernabéu, L. G. Cabral-Rosetti, J. Papavassiliou y J. Vidal 
\cite{bernabeu}. This form factor was obtained 
through the physical process $e^+ e^- \rightarrow \nu_l \bar{\nu_l}$ 
working in the linear $R_\xi$ gauge and using the electroweak BFM in the 
context of the standard model of the electroweak interactions, at the one 
loop level. We showed in frame of the SM, that the neutrino anapole form 
factor $F_{\chic A}(q^2)$ and the Dirac form factor $F_{\chic D}(q^2)$ are 
related as follows: $F_{\chic A}(q^2) = \frac{1}{q^2} F_{\chic D}(q^2)$. 
Therefore the neutrino anapole moment $a_{\nu_l}$ and the neutrino charge 
radius $\langle r^2_{\nu_l} \rangle$ satisfy the simple 

relation $a_{\nu_l} = \frac{1}{6}\langle r^2_{\nu_l} \rangle$. Hence, we 
showed that the anapole moment (as the charge radius) of the neutrino 
becomes a physical quantity, which has the following properties: (i)
it only gets contribution from the proper neutrino electromagnetic 
vertex, (ii) it is finite, (iii) it is independent on the gauge-fixing 
parameter, (iv) it is independent on the gauge-fixing scheme employed, 
(v) it does not depend on the Higgs or quark sector of the theory, (vi) 
it does not depend on the properties of the charged lepton used to define it. 
The numerical values of $a_{\nu_l}$ for the three different neutrinos are 
of the order $10^{-34} \, cm^2$.

\begin{theacknowledgments}
This work was supported in part by the {\it Programa de Apoyo a Proyectos de 
Investigación e Inovación Tecnológica} ({\bf PAPIIT}) de la 
{\bf DGAPA-UNAM} {\it No. de Proyecto}: {\sc IN109001}. 
\end{theacknowledgments}


\begin{thebibliography}{99}

\bibitem{abak} M. Abak and C. Aydin, {\it Europhys. Lett.} {\bf 4}, 881 (1987).

\bibitem{weinberg}
S.L. Glashow, {\it Nucl. Phys.} {\bf 22}, 579 (1961);\\
S. Weinberg, {\it Phys. Rev. Lett.} {\bf 19}, 1264 (1967);\\
A. Salam: {\it Proc. 8th Nobel Symposium}, p. 367, edited by N. Svartholm,
(Almqvist and Wiksell, Stockholm, 1968).

\bibitem{hooft} G. 't Hooft, {\it Nucl. Phys.} {\bf B35}, 167 (1971).

\bibitem{gongora} A. Góngora-T and Robin G. Stuart, {\it Z. Phys.} 
{\bf C55}, 101 (1992).

\bibitem{czyz} H. Czyz, K. Kolodziej, M. Zralek and P. Christova,
{\it Can. J. Phys.} {\bf 66}, 132 (1988);\\
H. Czyz and M. Zralek, {\it Can. J. Phys.} {\bf 66}, 384 (1988).

\bibitem{musolf} M. J. Musolf and B. R. Holstein, {\it Phys. Rev.} 
{\bf D43}, 2956 (1991).

\bibitem{rosado1} A. Rosado, {\it Phys. Rev.} {\bf D61}, 013001 (2000).

\bibitem{rosado2} A. Rosado, {\it Rev. Mex. Fís.} {\bf 47} (2), 132 (2001).

\bibitem{JMC} J. M. Cornwall, in {\it Proceeding of French-American Seminar
on Theoretical Aspects of Quamtun Chromodynamics}, Marseille, France, 1981,
edited by J. W. Dash (Centre de Physique Théorique, Marseille, 1982). 
J. M. Cornwall, {\it Phys. Rev. D} {\bf 26}, 1452 (1982).

\bibitem{CJM}J. M. Cornwall and J. Papavassiliou, {\it Phys. Rev.} {\bf D4}, 
3474 (1989).

\bibitem{JPP}J. Papavassiliou, {\it Phys. Rev.} {\bf D41}, 3179 (1990).

\bibitem{DAS}G. Degrassi and A. Sirlin, {\it Phys. Rev.} {\bf D46}, 
3104 (1992).

\bibitem{wat}N. J. Watson, To appear in the proceedings of the {\it 
International Workshop on Physical Variables in Gauge Theories}, Dubna, 
Russia, 21-25 Sep. 1999. , {\tt hep-ph/9912303}.

\bibitem{bernabeu} J. Bernabéu, L. G. Cabral-Rosetti, J. Papavassiliou
and J. Vidal, {\it Phys. Rev.} {\bf D62}, 113012 (2000).

\bibitem{chung} Chung Wook Kimm and Aihud Pevsher, {\it Neutrinos in Physics 
and Astrophysics}, Contemporary Concepts in Physics Volume 8, Harwood 
Academic Publishers 1993, in special see pp. 247-252.

\bibitem{mohapatra}
Rabindra N. Mohapatra and Palas B. Pal, {\it Massive Neutrinos in Physics
and Astrophysics}, Second Edition, World Scientific Lectures Notes in 
Physics Vol. 60, 1998, in especial see pp. 193-196.

\bibitem{cabral1} L. G. Cabral-Rosetti, J. Bernabéu, J. Vidal
and A. Zepeda, {\it Eur. Phys. J. C.} {\bf 12}, 633 (2000).

\bibitem{C3}L. G. Cabral-Rosetti, {\it Ph. D. Thesis} {\it ``Factores 
de Forma del Neutrino e Invariancia Gauge Electrodébil: El Radio de 
Carga''}, Departament de Física Te\`orica, Facultad de Fisiques, 
Universitat de Valencia, Estudi General, 11 de Diciembre de 2000, 
Valencia, Espa\~na.

\bibitem{BFM}A. Denner, G. Weiglein and S. Dittmaier {\it Nucl. Phys.} 
{\bf B440}, 95 (1995).

\bibitem{marshak} {Theory of weak interactions in particle physics}: R. E.
Marshak, Riazuddin and C. P. Ryan, Willey-Interscience 1998, p. 231.

\bibitem{cabral2} Luis G. Cabral-Rosetti and Miguel A. Sanchis-Lozano,
{\it J. Comp. Appl. Math.} {\bf 115}, 93 (2000).

\end{thebibliography}
\end{document}